\documentclass[twocolumn,preprintnumbers,amsmath,amssymb,superscriptaddress]{revtex4}
\usepackage{amsmath,amsfonts,amssymb,amsthm}  
\usepackage{graphicx}
\usepackage{graphicx}
\usepackage{color}

\begin{document}

\title{Multiparty quantum communication using multiqubit entanglement and teleportation}
\author {S. Ghose}
\affiliation{Department of Physics and Computer Science, Wilfrid Laurier University, Waterloo,  Canada}
\affiliation{Institute for Quantum Computing, University of Waterloo, Canada}
\email{sghose@wlu.ca}
\author{A. Kumar} 
\affiliation
{Indian Institute of Technology Rajasthan, Jodhpur, India}
\author{ V. Madhok}
\affiliation{Department of Physics and Computer Science, Wilfrid Laurier University, Waterloo,  Canada}

\author{A. M. Hamel}
\affiliation{Department of Physics and Computer Science, Wilfrid Laurier University, Waterloo, Canada}

\date{\today}

\begin{abstract}  
We propose a 2$N$ qubit entangled channel that can be used to teleport $N$ qubits in a network to a single receiver. We describe the construction of this channel and explicitly demonstrate how the protocol works. 
The protocol is different from teleportation using $N$ Bell pairs shared between $N$ senders and a single receiver, as it  ensures that all parties have to participate in order for the teleportation to be successful. This can be advantageous in various scenarios and we discuss
the potential application of this protocol to voting.
 \end{abstract}   
\maketitle

\section{Introduction}
Communication in a network of multiple parties is currently a vital area of research and discussion. 
Quantum protocols that harness the power of entanglement offer new avenues for efficient multipartite communication. For example, the pioneering protocol of Christandl and Wehner~\cite{Christandl} allows one member of a group to send another member anonymous messages.  However, their solution allows a malicious group member to alter the transmitted state undetected. Brassard et al.~\cite{Brassard} offer an improved anonymous communication protocol that sets no limit on the number of malicious participants and aborts if they are active. Identifying quantum  channels for  novel communication protocols thus remains an active area of research \cite{Muralidharan, Muralidharan1, Paul, Rigolin, Deng1, Man, Deng, Yeo, Chen, Man1, Wang, Yang}.
Such channels have potential applications to tasks such as voting, where private messages go to a single receiver. Quantum voting protocols have been proposed that use entangled states distributed amongst participants in order to allow each participant to act independently and in secret~\cite{Hillery, Vaccaro, Dolev, Horoshko, Bonanome, Li, Li1, Lang} .  Although the schemes ensure the anonymity of the senders, they require qubits/qudits to be physically sent to the receiver, which makes protocols vulnerable to eavesdropping or cheating.

In this paper we describe a novel protocol for communication using multiqubit entanglement and teleportation. We consider a network of $N$ parties, where each person wants to send a qubit privately and anonymously to a receiver. In the context of voting, each qubit could be used to encode a yes/no vote of the sender. One simple option for sending $N$ qubits to a receiver is for each member of the network to share an entangled Bell pair with the receiver,  and use this entangled channel to teleport his/her qubit to the receiver. Here, we consider a scenario that cannot be implemented with Bell pairs:  we look at the case where all members in the network must participate or the protocol fails  - i.e if any member does not send a qubit, then the receiver will not receive any of the qubits perfectly.  
Such an all-or-nothing scenario would be useful in certain voting schemes such as a jury or a committee, or in countries like Australia where all voters must participate. Furthermore, this protocol would be useful in cases where each transmitted qubit only encodes part of a message, and the message is readable only if all parts are teleported. For example, the qubits could encode parts of the password to a joint bank account shared by all $N$ parties that can only be accessed with the knowledge and participation of all members. Another possible application is in a spy network, where each spy in the network knows and transmits part of a secret. Unlike other secret sharing schemes \cite{Hillery1, Karlsson, Karlsson1, Gottesman, Tyc}, the senders do not have to communicate with each other and do not have any common information at the end of the protocol.  

We propose a 2$N$ qubit entangled channel that can be used to teleport N qubits to a receiver, where the nature of the entangled channel ensures that the protocol succeeds only when all $N$ senders participate. We show that  the the 2$N$ qubit entangled resource can be prepared from $N$ Bell pairs using $N-1$ CNOT gates. And unlike other schemes for $N$-qubit teleportation~\cite{Chen, Cheung, Muralidharan2, Zhao}, our protocol only requires single and two-qubit gates and measurements that are feasible in current experiments. Since the $N$ qubits are teleported to the receiver by the senders, there is no physical transmission of qubits from the senders to the receiver, and thus nothing for an adversary to steal or eavesdrop on. 
As we will show below, our protocol ensures that:\\
(1) Only authorized senders can teleport a qubit. \\
(2) A sender cannot transmit more than one qubit. In the context of voting, this means that each authorized voter can only vote once. \\
(3) No one can determine the state of anyone else's qubit (vote) from the publicly broadcast information in the protocol. \\
(4) The receiver will not be able to gain any information about the individual qubits from the publicly broadcast information in the protocol. The receiver does not need to know the state of each qubit to accurately receive each qubit.  \\
(5) A sender cannot prove to a third party what his/her teleported state was. For voting, this means that the vote is receipt free - i.e -the voter cannot prove to a third party how he/she voted.\\
In order to ensure these conditions, our study requires some assumptions to be made: \\
1. The senders and the receiver verify that they share a perfect entangled channel using established techniques. 
Once the senders and the receiver verify that they share a perfect quantum channel, the protocol ensures that only authorized senders participate in the protocol. \\
2. All the senders and the receiver in the protocol are honest, i.e. none of them tries to cheat during the protocol.    \\
Under these assumptions, our protocol ensures teleportation of $N$ independent qubits to the receiver with perfect fidelity and unit probability when all $N$ parties participate. We now proceed to discuss the protocol in detail. \par

\section{$2N$-qubit entangled quantum channel}
In order to facilitate the discussion of our $2N$ qubit entangled channel and the results obtained in this article, we first briefly describe the properties of a quantum channel composed of the direct product of $N$ Bell pairs $\left|\phi\right\rangle^{+}=\frac{1}{\sqrt{2}}\left[\left|00\right\rangle+\left|11\right\rangle\right]$, i.e. 
\begin{eqnarray}
\left|\Phi\right\rangle_{12.....(2N - 1)(2N)} &=& \left|\phi\right\rangle_{(1)(N+1)}^{+} \otimes \left|\phi\right\rangle_{(2)(N+2)}^{+} \hdots \nonumber \\ &\hdots& \otimes \left|\phi\right\rangle_{(N-1)(2N-1)}^{+} \otimes \left|\phi\right\rangle_{(N)(2N)}^{+} \nonumber \\ &&
\end{eqnarray}
where the $1^{st}$ qubit is entangled with the $(N+1)$-th qubit, $2^{nd}$ qubit is entangled with the $(N+2)$-th qubit and so on. The $2N$ qubit channel in Eq. (1) has the property that the value (0 or 1) of the first $N$ qubits is exactly the same as that of the last $N$ qubits. For example, the four and six-qubit quantum channels corresponding to the direct product of two and three Bell pairs can be written as    
\begin{eqnarray}
\left| \Phi  \right\rangle _{1234}  &=& \left|\phi\right\rangle^{+}_{13} \otimes \left|\phi\right\rangle^{+}_{24}  \nonumber \\ &=& \frac{1}{\sqrt{2}}\left[\left|00 \right\rangle  + \left| 11 \right\rangle\right]_{13} \otimes \frac{1}{\sqrt{2}}\left[\left|00 \right\rangle  + \left| 11 \right\rangle\right]_{24} \nonumber \\ &=&
\frac{1}{2}\left[\left|0000 \right\rangle  + \left| 0101 \right\rangle  + \left| 1010 \right\rangle+\left| 1111 \right\rangle\right]_{1234} \nonumber \\
\end{eqnarray}
and
\begin{eqnarray}
\left| \Phi  \right\rangle _{123456}  &=& \left|\phi\right\rangle^{+}_{14} \otimes \left|\phi\right\rangle^{+}_{25} \otimes \left|\phi\right\rangle^{+}_{36} \nonumber \\ &=& \frac{1}{2}\left[\left|0000 \right\rangle  + \left| 0101 \right\rangle  + \left| 1010 \right\rangle+\left| 1111 \right\rangle\right]_{1245} \nonumber \\ &\otimes& \frac{1}{\sqrt{2}}\left[\left|00 \right\rangle  + \left| 11 \right\rangle\right]_{36} \nonumber \\ &=&
\frac{1}{2\sqrt{2}}\left[\left|000000 \right\rangle  + \left| 001001 \right\rangle  + \left| 010010 \right\rangle\right. \nonumber \\  &+& \left.\left| 011011 \right\rangle+ \left| 100100 \right\rangle  + \left| 101101 \right\rangle \right. \nonumber \\  &+& \left. \left| 110110 \right\rangle  + \left| 111111 \right\rangle\right]_{123456}\nonumber\\
\end{eqnarray}
respectively. 
Notice that the first 2 qubits in each term in Eq. (2)  are identical to the last 2 qubits, and the first 3 qubits in Eq. (3) are identical to the last 3 qubits in each term. This property allows us to analyze the characteristics of the $2N$ qubit entangled quantum state used as a resource in this article.  \par
The $2N$ qubit entangled channel $\left|\Psi\right\rangle_{12.....(2N - 1)(2N)}$ can be obtained from the direct product state $\left| \Phi  \right\rangle _{12...N(N + 1)..(2N - 1)(2N)}$ in Eq. (1) by performing $(N-1)$ CNOT operations. This can be achieved by using the $2N$-th qubit as target and the qubits $1$, $2$, ...... $(N-2)$, $(N-1)$ 
 as controls.
For each term of the $2N$ qubit entangled state $\left|\Psi\right\rangle_{12.....(2N - 1)(2N)}$, the value (0 or 1) of the $2N$-th qubit is exactly the same as that of the $N$-th qubit if the number of $\left|1\right\rangle$'s from the qubits $1$, $2$, ......, $(N-1)$     is even, else the value of the $2N$-th qubit is opposite to that of the $N$-th qubit. Hence, the $2N$ qubit quantum channel can be written as 
\begin{widetext}
\begin{eqnarray}
\lefteqn{\left|\Psi\right\rangle _{12...(2N - 1)2N} =} && \nonumber \\ && \frac{1}{\sqrt{2^{N-1}}}\left[\sum\limits_{i}\{\left|\psi_{i} \right\rangle_{1...(N - 1)}\left|\psi_{i} \right\rangle_{(N+1)...(2N-1)}\}\left|\phi\right\rangle_{N(2N)}^{+} +\sum\limits_{j}\{\left|\psi_{j} \right\rangle_{1...(N - 1)}\left|\psi_{j}\right\rangle_{(N+1)...(2N-1)}\}\left|\psi\right\rangle_{N(2N)}^{+}\right] \nonumber \\ &&
\end{eqnarray}  
\end{widetext}
where the state $\left|\psi_{i}\right\rangle$ contains half of the basis states corresponding to $(N-1)$ qubits with even number of $\left|1\right\rangle$'s and  the state $\left|\psi_{j}\right\rangle$ contains other half of the basis states with odd number of $\left|1\right\rangle$'s. For example, the states $\left|\psi_{i}\right\rangle$ and $\left|\psi_{j}\right\rangle$ for a four-qubit quantum channel would be 
$\left|\psi_{i}\right\rangle_{1}=\left|\psi_{i}\right\rangle_{3}=\left|0\right\rangle$ and  $\left|\psi_{j}\right\rangle_{1}=\left|\psi_{j}\right\rangle_{3}=\left|1\right\rangle$. Similarly for a six-qubit quantum channel, the states $\left|\psi_{i}\right\rangle$ and $\left|\psi_{j}\right\rangle$ are
$\left|\psi_{i}\right\rangle_{12}=\left|\psi_{i}\right\rangle_{45}=\left|00\right\rangle, \left|11\right\rangle$ and $\left|\psi_{j}\right\rangle_{12}=\left|\psi_{j}\right\rangle_{45}=\left|01\right\rangle, \left|10\right\rangle$, respectively.  Hence, the four and six-qubit entangled channels would be
\begin{widetext}
\begin{eqnarray}
\left| \Psi  \right\rangle _{1234}  &=& \frac{1}{2}\left[\left|00\right\rangle_{13}  \otimes \frac{1}{\sqrt{2}}\left\{ {\left| {00} \right\rangle  + \left| {11} \right\rangle } \right\}_{24}  + \left|11\right\rangle_{13}  \otimes \frac{1}{{\sqrt 2 }}\left\{ {\left| {01} \right\rangle  + \left| {10} \right\rangle } \right\}_{24}  \right] \nonumber \\ 
&=& \frac{1}{2}\left[\left|0000 \right\rangle  + \left| 0101 \right\rangle  + \left| 1011 \right\rangle+\left| 1110 \right\rangle\right]_{1234}\nonumber\\
\left| \Psi  \right\rangle _{123456}  &=& \frac{1}{2}\left[ {\left\{ {\left| {0000} \right\rangle  + \left| {1111} \right\rangle } \right\}_{1245}  \otimes \frac{1}{{\sqrt 2 }}\left\{ {\left| {00} \right\rangle  + \left| {11} \right\rangle } \right\}_{36}  + \left\{ {\left| {0101} \right\rangle  + \left| {1010} \right\rangle } \right\}_{1245}  \otimes \frac{1}{{\sqrt 2 }}\left\{ {\left| {01} \right\rangle  + \left| {10} \right\rangle } \right\}_{36} } \right] \nonumber \\ 
&=& \frac{1}{2\sqrt{2}}\left[\left|000000 \right\rangle  + \left| 001001 \right\rangle  + \left| 010011 \right\rangle+\left| 011010 \right\rangle+ \left| 100101 \right\rangle  + \left| 101100 \right\rangle + \left| 110110 \right\rangle  + \left| 111111 \right\rangle\right]_{123456}, \nonumber \\ &&
\end{eqnarray}
\end{widetext}

Notice that in each term, the fourth qubit (sixth qubit) in Eq. (5) (Eq.(6)) is exactly the same as the second qubit (third qubit) if the number of $\left|1\right\rangle$'s in the individual terms $\left|0\right\rangle$, $\left|1\right\rangle$ $(\left|00\right\rangle$, $\left|01\right\rangle$, $\left|10\right\rangle$, $\left|11\right\rangle)$ of the state $\left|\psi\right\rangle_{1}/_{3}$ $(\left|\psi\right\rangle_{12}/_{45})$ is even, else the the sixth (eighth) qubit is the opposite of the third (fourth) qubit.

\section {N-qubit communication protocol}
We first review the original teleportation protocol proposed by Bennett {\em et al} \cite{Bennett} in order to provide an insight into our scheme. If a sender wants to communicate a single qubit state $\left| \phi  \right\rangle_{1}  = \left[a\left| 0 \right\rangle + b\left| 1  \right\rangle\right]_{1}$ to a receiver, then the sender and the receiver must share one of the Bell states, i.e. 
\begin{eqnarray}
\left| \psi  \right\rangle _{23}^ \pm   &=& \frac{1}{{\sqrt 2 }}\left[ {\left| {01} \right\rangle  \pm \left| {10} \right\rangle } \right]_{23}, \nonumber \\ 
\left| \phi  \right\rangle _{23}^ \pm   &=& \frac{1}{{\sqrt 2 }}\left[ {\left| {00} \right\rangle  \pm \left| {11} \right\rangle } \right]_{23} 
\end{eqnarray}
where the qubit 2 is assigned to the sender and the qubit 3 is assigned to the receiver. If the shared entangled resource is $\left| \phi  \right\rangle _{23}^{+}$ then the joint state of the three qubits can be written as
\begin{eqnarray}
\left| \psi  \right\rangle _{234}  &=& \left| \phi  \right\rangle_1  \otimes \left| \phi  \right\rangle _{23}^{+}  \nonumber \\ 
&=& \frac{{\left| \phi  \right\rangle _{12}^ +  }}{{2}}\left[ a\left| {0} \right\rangle  + b\left| {1} \right\rangle\right]_{3}+\frac{{\left| \phi  \right\rangle _{12}^ -  }}{{2}}\left[ a\left| {0} \right\rangle  - b\left| {1} \right\rangle\right]_{3}  \nonumber \\ 
&+& \frac{{\left| \psi  \right\rangle _{12}^ +  }}{{2}}\left[ a\left| {1} \right\rangle  + b\left| {0} \right\rangle\right]_{3}+\frac{{\left| \psi  \right\rangle _{12}^ -  }}{{2}}\left[ a\left| {1} \right\rangle -  b\left| {0} \right\rangle\right]_{3} \nonumber \\ &&
\end{eqnarray}
If the sender performs any measurement on the qubits 1 and 2 in the Bell basis, then the state of receiver's qubit will be projected onto one of the four possible states as shown in Eq. (8) with equal probability of $1/4$. For example, if the sender's measurement outcome is $\left|\phi\right\rangle_{12}^{+}$, then the receiver's qubit will be projected onto the desired teleported state. However, in all the other measurement outcomes of the sender, the receiver would need to perform a single qubit unitary transformation on the qubit 3 to recover the teleported state.  \par
The direct product state of $N$ Bell pairs $\left|\phi\right\rangle_{ij}^{+}$ in Eq. (1) can also be used as a resource for the teleportation of $N$ single qubit states in a similar fashion. For example, if the $2N$ qubit channel in Eq. (1) is shared between the $N$ senders and a receiver, then the first $N$ qubits can be assigned to the $N$ senders and the last $N$ qubits can be assigned to the receiver. Each sender wants to teleport a single-qubit state to the receiver and thus measures his/her qubits in the Bell basis and sends the measurement result to the receiver. The receiver would then be able to recover the state of all the qubits by performing single-qubit unitary transformations on his/her qubits depending on the measurement outcomes of the senders. Notice that similar to the single qubit teleportation protocol, the measurement outcomes $\left|\phi\right\rangle_{ij}^{+}$, $\left|\phi\right\rangle_{ij}^{-}$, $\left|\psi\right\rangle_{ij}^{+}$ and $\left|\psi\right\rangle_{ij}^{-}$ would always correspond to the states $\left[a\left| {0} \right\rangle  + b\left| {1} \right\rangle\right]$, $\left[a\left| {0} \right\rangle  - b\left| {1} \right\rangle\right]$, $\left[a\left| {1} \right\rangle  + b\left| {0} \right\rangle\right]$ and $\left[a\left| {1} \right\rangle - b\left| {0} \right\rangle\right]$, respectively. \par

We now proceed to describe the teleportation protocol using our state as an entangled channel. 
For our protocol, the first $N$ qubits in the shared entangled state are with the senders and the last $N$ qubits are with the receiver. Each sender wants to teleport the state $\left| \phi  \right\rangle_{A_{i}}  = \left[a_{A_{i}} \left| 0 \right\rangle_{A_{i}}   + b_{A_{i}} \left| 1_{A_{i}}  \right\rangle\right]$ to the receiver. Thus, the joint state of $3N$ qubits ,
composed of $N$ votes and $2N$ qubit entangled channel can be expressed as
\begin{eqnarray}
\lefteqn{\left| \psi  \right\rangle _{12...(3N-1)3N} =} && \nonumber \\ &&   \prod\limits_i {\left[ {a_{A_{i}} \left| 0 \right\rangle_{A_{i}}  + b_{A_{i}} \left| 1 \right\rangle_{A_{i}} } \right] }  \otimes \left| \psi  \right\rangle _{12...(2N-1)2N} 
\end{eqnarray}
Each sender performs a Bell measurement on his/her qubit to be teleported,  and the shared entangled qubit in the $2N$ qubit entangled state. The joint state of $3N$ qubits in Eq. (9) can be rewritten in terms of the measurement basis of all the senders as 
\begin{widetext}
\begin{eqnarray}
\left| \psi  \right\rangle  = \frac{1}{{2^N }}\left[ \begin{array}{l}
 \left\{ {\left| \phi  \right\rangle _{A_1 1}^ +   \otimes \left| \phi  \right\rangle _{A_2 2}^ +   \otimes ...... \otimes \left| \phi  \right\rangle _{A_{N - 1} (N - 1)}^ +   \otimes \left| \phi  \right\rangle _{A_N N}^ +  } \right\}\left| \chi  \right\rangle _{(N + 1)(N + 2)......(2N - 1)2N}^{(1)}  \\ 
  + \left\{ {\left| \phi  \right\rangle _{A_1 1}^ +   \otimes \left| \phi  \right\rangle _{A_2 2}^ +   \otimes ...... \otimes \left| \phi  \right\rangle _{A_{N - 1} (N - 1)}^ +   \otimes \left| \phi  \right\rangle _{A_N N}^ -  } \right\}\left| \chi  \right\rangle _{(N + 1)(N + 2)......(2N - 1)2N}^{(2)}  \\ 
  +  \hdots\\ 
 \, \vdots  \\ 
 \, \vdots  \\ 
 + \hdots \\
  + \left\{ {\left| \psi  \right\rangle _{A_1 1}^ -   \otimes \left| \psi  \right\rangle _{A_2 2}^ -   \otimes ...... \otimes \left| \psi  \right\rangle _{A_{N - 1} (N - 1)}^ -   \otimes \left| \psi  \right\rangle _{A_N N}^ +  } \right\}\left| \chi  \right\rangle _{(N + 1)(N + 2)......(2N - 1)2N}^{(4^N - 1)}  \\ 
  + \left\{ {\left| \psi  \right\rangle _{A_1 1}^ -   \otimes \left| \psi  \right\rangle _{A_2 2}^ -   \otimes ...... \otimes \left| \psi  \right\rangle _{A_{N - 1} (N - 1)}^ -   \otimes \left| \psi  \right\rangle _{A_N N}^ -  } \right\}\left| \chi  \right\rangle _{(N + 1)(N + 2)......(2N - 1)2N}^{(4^N)}  \\ 
 \end{array} \right]
\end{eqnarray}
\end{widetext}
where $\left| \chi  \right\rangle _{(N + 1)(N + 2)......(2N - 1)2N}^{(1)}$ $-$ $\left| \chi  \right\rangle _{(N + 1)(N + 2)......(2N - 1)2N}^{(4^N)}$ are the states of receiver's qubits depending on the measurement outcomes of all the senders and contains all the information about the individual qubits (votes). For example, if we consider a four-qubit entangled resource, then the $\left|\chi\right\rangle_{34}^{(1)}$ state corresponding to the measurement outcomes $\left|\phi\right\rangle_{A_{1}1}^{+}$ and $\left|\phi\right\rangle_{A_{2}2}^{+}$ would be
\begin{widetext}
\begin{eqnarray}
\left| \chi  \right\rangle _{34}^{(1)}  &=& \left[a_{A_1} a_{A_2 }  \left| {00} \right\rangle _{34}  + a_{A_1 } b_{A_2 } \left| {01} \right\rangle _{34} + b_{A_1 } a_{A_2 } \left| {11} \right\rangle _{34} + b_{A_1 } b_{A_2 } \left| {10} \right\rangle _{34} \right]
\end{eqnarray}
\end{widetext}
\begin{table*}
\caption{\label{tab:table3} 2-qubit teleportation: unitary transformations required for the receiver to recover two qubits depending on the measurement outcomes of the two senders.}
\begin{ruledtabular}
\begin{tabular}{cccc}
First sender's & Second sender's & Final state with the receiver & Unitary transformations required to  \\  measurements &  measurements & after the CNOT operation &   obtain the correct votes corresponding to the  \\ & & & measurement outcomes $(++$, $+-$, $-+$, $--)$ \\ \hline
 $\left|\phi\right\rangle_{A_{1}1}^{\pm}$&$\left|\phi\right\rangle_{A_{2}2}^{\pm}$&$\left[a_{A_{1}1}\left|0\right\rangle_{3} \pm b_{A_{1}1}\left|1\right\rangle_{3}\right]\otimes\left[a_{A_{2}2}\left|0\right\rangle_{3} \pm b_{A_{2}2}\left|1\right\rangle_{4}\right]$ & $I$, $\sigma_{z}^{4}$, $\sigma_{z}^{3}$,  $\sigma_{z}^{3}\sigma_{z}^{4}$  \\ & & & \\
 $\left|\phi\right\rangle_{A_{1}1}^{\pm}$&$\left|\psi\right\rangle_{A_{2}2}^{\pm}$
 &$\left[a_{A_{1}1}\left|0\right\rangle_{3} \pm b_{A_{1}1}\left|1\right\rangle_{3}\right]\otimes\left[a_{A_{2}2}\left|1\right\rangle_{3} \pm b_{A_{2}2}\left|0\right\rangle_{4}\right]$ & $\sigma_{x}^{4}$, $\sigma_{y}^{4}$, $\sigma_{z}^{3}\sigma_{x}^{4}$, $\sigma_{z}^{3}\sigma_{y}^{4}$ \\ & & & \\
 $\left|\psi\right\rangle_{A_{1}1}^{\pm}$&$\left|\phi\right\rangle_{A_{2}2}^{\pm}$&$\left[a_{A_{1}1}\left|1\right\rangle_{3} \pm b_{A_{1}1}\left|0\right\rangle_{3}\right]\otimes\left[a_{A_{2}2}\left|0\right\rangle_{3} \pm b_{A_{2}2}\left|1\right\rangle_{4}\right]$ & $\sigma_{x}^{3}$, $\sigma_{x}^{3}\sigma_{z}^{4}$, $\sigma_{y}^{3}$, $\sigma_{y}^{3}\sigma_{z}^{4}$ \\ & & & \\
 $\left|\psi\right\rangle_{A_{1}1}^{\pm}$&$\left|\psi\right\rangle_{A_{2}2}^{\pm}$&$\left[a_{A_{1}1}\left|1\right\rangle_{3} \pm b_{A_{1}1}\left|0\right\rangle_{3}\right]\otimes\left[a_{A_{2}2}\left|1\right\rangle_{3} \pm b_{A_{2}2}\left|0\right\rangle_{4}\right]$ & $\sigma_{x}^{3}\sigma_{x}^{4}$, $\sigma_{x}^{3}\sigma_{y}^{4}$, $\sigma_{y}^{3}\sigma_{x}^{4}$, $\sigma_{y}^{3}\sigma_{y}^{4}$ \\
\end{tabular}
\end{ruledtabular}
\end{table*}


 The receiver at this stage would not be able to separate the exact qubits teleported sent by the senders as the joint state of receiver's qubit cannot be written as the direct product state of the individual qubits. However, if he/she performs a CNOT operation with qubit 3 as the control and qubit 4 as the target, then the joint state of the receiver's qubits can be re-expressed as 
\begin{widetext}
\begin{eqnarray}
\left| \chi  \right\rangle _{34}^{(1)}  &=& \left[a_{A_1} a_{A_2 }  \left| {00} \right\rangle _{34}  + a_{A_1 } b_{A_2 } \left| {01} \right\rangle _{34} + b_{A_1 } a_{A_2 } \left| {10} \right\rangle _{34} + b_{A_1 } b_{A_2 } \left| {11} \right\rangle _{34} \right] \nonumber \\
&=& \left[a_{A_{1}} \left| 0 \right\rangle_{3}   + b_{A_{1}} \left| 1_{3}  \right\rangle\right] \otimes \left[a_{A_{2}} \left| 0 \right\rangle_{4}   + b_{A_{2}} \left| 1_{4}  \right\rangle\right]
\end{eqnarray}
\end{widetext}
respectively. Thus, the receiver can successfully separate the qubits teleported by the senders. 
However, for all the other measurement outcomes of the senders, the receiver would need to perform single qubit unitary operations in addition to the CNOT transformations.  To see why the protocol would work for all possible measurement outcomes, we can use the principle of `deferred measurement'.  By the principle of `deferred measurement' it is  possible to defer all the measurements until after the receiver's CNOT operations. Using this, our protocol can be shown to be equivalent to the standard teleportation using $N$ Bell pairs as the CNOT transformations applied to create the channel will cancel the CNOT transformations required at the later stage by the receiver to separate the qubits. \par
The required single-qubit unitary transformations for the use of the four qubit entangled channel are summarized in Table 1. For this protocol to be successful, the receiver must
know the distribution of entangled qubits so that he/she can apply the correct unitary transformations to recover
the teleported qubits accurately. However, this does not allow the receiver to gain any information about the individual quibits. Hence, in the context of voting, where each teleported qubit encodes a vote, the receiver would always be able to tally the correct votes sent by the senders but would not be able to gain perfect information about all the individual votes. The multiqubit state teleported to the receiver could also be used as an input for another computation or communication protocol without the receiver having to know the state of the qubits.

For the case of a $2N$ qubit entangled channel, the state $\left|\chi\right\rangle_{(N+1)(N+2)...(2N-1)2N}^{(1)}$ corresponding to the measurement outcomes $\left|\phi\right\rangle_{A_{1}1}^{+}$, $\left|\phi\right\rangle_{A_{2}2}^{+}$$\hdots$$\hdots$, $\left|\phi\right\rangle_{A_{N-1}(N-1)}^{+}$ and $\left|\phi\right\rangle_{A_{N}(N)}^{+}$ can be either written as
\begin{widetext}
\begin{eqnarray}
\begin{array}{l}
 \left| \chi  \right\rangle _{(N + 1)(N + 2)...(2N - 1)2N}^{(1)}  \\ 
  = \left[ \begin{array}{l}
 a_{A_1 } a_{A_2 } ...a_{A_{N - 1} } a_{A_N } \left| {00...00} \right\rangle _{(N + 1)(N + 2)...(2N - 1)2N}  + a_{A_1 } a_{A_2 } ...a_{A_{N - 1} } b_{A_N } \left| {00...01} \right\rangle _{(N + 1)(N + 2)...(2N - 1)2N}  \\ 
  + a_{A_1 } a_{A_2 } ...b_{A_{N - 1} } a_{A_N } \left| {00...11} \right\rangle _{(N + 1)(N + 2)...(2N - 1)2N}  + a_{A_1 } a_{A_2 } ...b_{A_{N - 1} } b_{A_N } \left| {00...10} \right\rangle _{(N + 1)(N + 2)...(2N - 1)2N}  \\ 
  +  \\ 
 \, \vdots  \\ 
  + b_{A_1 } b_{A_2 } ...a_{A_{N - 1} } a_{A_N } \left| {11...00} \right\rangle _{(N + 1)(N + 2)...(2N - 1)2N}  + b_{A_1 } b_{A_2 } ...a_{A_{N - 1} } b_{A_N } \left| {11...01} \right\rangle _{(N + 1)(N + 2)...(2N - 1)2N}  \\ 
  + b_{A_1 } b_{A_2 } ...b_{A_{N - 1} } a_{A_N } \left| {11...11} \right\rangle _{(N + 1)(N + 2)...(2N - 1)2N}  + b_{A_1 } b_{A_2 } ...b_{A_{N - 1} } b_{A_N } \left| {11...10} \right\rangle _{(N + 1)(N + 2)...(2N - 1)2N}  \\ 
 \end{array} \right] \\ 
 \end{array}
\end{eqnarray}
\end{widetext}
if $N$ is even or
\begin{widetext}
\begin{eqnarray}
\begin{array}{l}
 \left| \chi  \right\rangle _{(N + 1)(N + 2)...(2N - 1)2N}^{(1)}  \\ 
  = \left[ \begin{array}{l}
 a_{A_1 } a_{A_2 } ...a_{A_{N - 1} } a_{A_N } \left| {00...00} \right\rangle _{(N + 1)(N + 2)...(2N - 1)2N}  + a_{A_1 } a_{A_2 } ...a_{A_{N - 1} } b_{A_N } \left| {00...01} \right\rangle _{(N + 1)(N + 2)...(2N - 1)2N}  \\ 
  + a_{A_1 } a_{A_2 } ...b_{A_{N - 1} } a_{A_N } \left| {00...11} \right\rangle _{(N + 1)(N + 2)...(2N - 1)2N}  + a_{A_1 } a_{A_2 } ...b_{A_{N - 1} } b_{A_N } \left| {00...10} \right\rangle _{(N + 1)(N + 2)...(2N - 1)2N}  \\ 
  +  \\ 
 \, \vdots  \\ 
  + b_{A_1 } b_{A_2 } ...a_{A_{N - 1} } a_{A_N } \left| {11...01} \right\rangle _{(N + 1)(N + 2)...(2N - 1)2N}  + b_{A_1 } b_{A_2 } ...a_{A_{N - 1} } b_{A_N } \left| {11...00} \right\rangle _{(N + 1)(N + 2)...(2N - 1)2N}  \\ 
  + b_{A_1 } b_{A_2 } ...b_{A_{N - 1} } a_{A_N } \left| {11...10} \right\rangle _{(N + 1)(N + 2)...(2N - 1)2N}  + b_{A_1 } b_{A_2 } ...b_{A_{N - 1} } b_{A_N } \left| {11...11} \right\rangle _{(N + 1)(N + 2)...(2N - 1)2N}  \\ 
 \end{array} \right] \\ 
 \end{array}
\end{eqnarray}
\end{widetext}
if $N$ is odd. 
In order to recover the teleported qubits, the receiver performs $(N-1)$ CNOT operations on the joint state of $N$ qubits by using the $2N$-th qubit as the target and the qubits $(N+1)$, $(N+1)$ ,......, $(2N-2)$, $(2N-1)$ as controls, respectively. The CNOT operations transform the states in Eq. (16) and (17) to the form
\begin{eqnarray}
\left| \chi  \right\rangle _{(N + 1)(N + 2)...(2N - 1)2N}^{(1)}  &=& \prod\limits_i {\left( {a_{A_i } \left| 0 \right\rangle _{N + i}  + b_{A_i } \left| 1 \right\rangle _{N + i} } \right)} \nonumber \\ &&
\end{eqnarray}
In this specific case, the receiver does not need to perform any single-qubit unitary transformations. However, for all other sets of measurement outcomes, the receiver would need to apply single-qubit transformations to his/her qubits conditioned on the senders' measurement outcomes to exactly recover the teleported qubits. Notice that the measurement outcomes $\left|\phi\right\rangle_{A_i i}^ +$, $\left|\phi\right\rangle_{A_i i}^ -$, $\left|\psi\right\rangle_{A_i i}^ +$ and $\left|\psi\right\rangle_{A_i i}^ -$ correspond to the states $\left[ {a_{A_i } \left| 0 \right\rangle _{N + i}  + b_{A_i } \left| 1 \right\rangle _{N + i} } \right]$, $\left[ {a_{A_i } \left| 0 \right\rangle _{N + i}  - b_{A_i } \left| 1 \right\rangle _{N + i} } \right]$, $\left[ {a_{A_i } \left| 1 \right\rangle _{N + i}  + b_{A_i } \left| 0 \right\rangle _{N + i} } \right]$ and $\left[ {a_{A_i } \left| 1 \right\rangle _{N + i}  - b_{A_i } \left| 0 \right\rangle _{N + i} } \right]$, respectively. For example, if the measurement outcomes of all the senders are $\left|\psi\right\rangle_{A_{i}i}^{-}$ then in addition to the $(N-1)$ CNOT operations, the receiver needs to apply $\sigma_{y}^{i}$ operations to all the qubits. 
Similarly, for all other sets of measurement outcomes, the receiver has to apply the appropriate unitary to each qubit in order to recover 
all the teleported qubits accurately. 

Our protocol requires all senders to participate for the teleportation to be successful. Once all the senders broadcast their results, the receiver performs all the CNOT operations.
If one or more senders decides not to participate in the scheme and does not measure his/her qubit, then the receiver does not perform any of the CNOT operations. Note that all the receiver's CNOT operations use the $2N$th qubit as a target. Thus, in order to ensure that the receiver only performs all the CNOT operations when all the senders participate, one could require that the receiver only gets access to the $2N$th qubit after all the senders participate and broadcast their measurement results. In comparison, for the standard teleportation using $N$ Bell pairs, limiting the receiver's access to only the $2N$th qubit would not prevent the successful teleportation of the 1st $N-1$ qubits. Our protocol thus offers the possibility of a scheme that cannot be done with $N$ Bell pairs. Securing access to the $2N$th qubit also has an additional benefit: if the other qubits are compromised or stolen, the teleported state of the qubits cannot be recovered from the stolen qubits, since the $2N$th qubit is required as a target to perform the CNOT operations and recover the teleported state of the other qubits. Thus, by keeping the $2N$th qubit secure, the protocol's security is ensured.
In comparison, for the standard teleportation using $N$ Bell pairs, all the qubits have to be kept safe in order to ensure secure teleportation. In a network as the number of senders, $N$ grows larger, the benefit of our protocol increases.

Previous work has identified some general rules that should be satisfied by voting protocols~\cite{Vaccaro}. We describe below how our protocol ensures these rules  are met:\\
(1) Only authorized voters can vote.  This is ensured by the fact that the entangled channel qubits are only distributed to authorized voters. Without sharing a qubit from the entangled channel, a voter will not be able to transmit his/her vote.\\
(2) Each authorized voter can only vote once. Notice that once a sender performs a Bell measurement on his/her qubit from the entangled channel, it becomes disentangled from
the rest of the system. Thus, the sender cannot teleport additional qubits, which ensures that a sender in a voting scheme cannot vote more than once. \\
(3) No one can determine the state of anyone else's vote from the publicly broadcast information in the protocol. Since the only publicly broadcast messages are the measurement outcomes of each voter, the individual votes cannot be deterministically extracted from this public information.\\
(4) The receiver will also not be able to gain any information about the individual votes from the publicly broadcast measurement outcomes. The receiver does not need to know the state of each qubit to accurately receive each qubit. If the receiver wants to know what is the yes/no vote of each voter, he/she would have to additionally measure each qubit in the basis in which the votes were encoded. Thus, in order to prevent the receiver from getting to know the votes, the encoding basis could be kept secret from the receiver.  \\
(5) Each vote is receipt free. Each voter's Bell measurement destroys the original state that was to be teleported, so a voter cannot show his/her vote to a third party after the vote has been teleported. Furthermore, the voter cannot use the measurement outcomes to prove to a third party what his/her vote was.\\
Our protocol thus allows the successful implementation of a voting scheme that satisfies these general rules.

\section{Conclusion}

The development of resources for quantum communication and computation lies at the heart of quantum information theory. 
We have described a $2N$ qubit entangled channel that can be used to teleport $N$ independent qubits in a network to a single receiver. We have discussed the construction of this channel and demonstrated how the protocol can be used for teleporting $N$ qubits using only single and 2-qubit operations. We have shown the explicit operations required for 2 and 3-qubit teleportation and described the general $N$-qubit teleportation scheme. 
Our protocol is different from teleportation of qubits using a product of $N$ Bell states. The addition of  CNOT gates to the standard teleportation protocol via $N$ Bell pairs makes it possible to design new communication scenarios that cannot be implemented with $N$ Bell states. The structure of the channel constructed ensures that all authorized parties  that have a qubit from the entangled channel have to participate in the protocol. This can be advantageous in situations where we demand all parties must participate. As an example, we have discussed the potential application of this protocol in a voting scheme. This scheme ensures that all voters must participate, that only authorized voters can vote, that each voter can only vote once and that the vote is receipt-free. Unlike other quantum voting schemes,
a third party cannot intercept or steal the votes. since the votes are teleported rather than physically sent. Assuming that the locations of the voters are isolated from each other, then the votes are private since the individual votes cannot be extracted perfectly from the publicly broadcast measurement outcomes. Our results thus demonstrate the potential of using multiqubit entanglement for communication tasks tailored to specific needs. We hope to extend this idea to develop other novel schemes in the future.

\section{Acknowledgements}
SG and AMH were each supported by an NSERC Discovery Grant. VM acknowledges funding from the Ontario Ministry of Research and Innovation and Wilfrid Laurier University.

\section{Appendix}

We explicitly demonstrate our protocol for the case of 3-qubit teleportation, using a six-qubit entangled channel.
The joint state of $3N$ qubits in Eq. (9) can be rewritten in terms of the measurement basis of all the senders as 
\begin{widetext}
\begin{eqnarray}
\left| \psi  \right\rangle  = \frac{1}{{2^N }}\left[ \begin{array}{l}
 \left\{ {\left| \phi  \right\rangle _{A_1 1}^ +   \otimes \left| \phi  \right\rangle _{A_2 2}^ +   \otimes ...... \otimes \left| \phi  \right\rangle _{A_{N - 1} (N - 1)}^ +   \otimes \left| \phi  \right\rangle _{A_N N}^ +  } \right\}\left| \chi  \right\rangle _{(N + 1)(N + 2)......(2N - 1)2N}^{(1)}  \\ 
  + \left\{ {\left| \phi  \right\rangle _{A_1 1}^ +   \otimes \left| \phi  \right\rangle _{A_2 2}^ +   \otimes ...... \otimes \left| \phi  \right\rangle _{A_{N - 1} (N - 1)}^ +   \otimes \left| \phi  \right\rangle _{A_N N}^ -  } \right\}\left| \chi  \right\rangle _{(N + 1)(N + 2)......(2N - 1)2N}^{(2)}  \\ 
  +  \hdots\\ 
 \, \vdots  \\ 
 \, \vdots  \\ 
 + \hdots \\
  + \left\{ {\left| \psi  \right\rangle _{A_1 1}^ -   \otimes \left| \psi  \right\rangle _{A_2 2}^ -   \otimes ...... \otimes \left| \psi  \right\rangle _{A_{N - 1} (N - 1)}^ -   \otimes \left| \psi  \right\rangle _{A_N N}^ +  } \right\}\left| \chi  \right\rangle _{(N + 1)(N + 2)......(2N - 1)2N}^{(4^N - 1)}  \\ 
  + \left\{ {\left| \psi  \right\rangle _{A_1 1}^ -   \otimes \left| \psi  \right\rangle _{A_2 2}^ -   \otimes ...... \otimes \left| \psi  \right\rangle _{A_{N - 1} (N - 1)}^ -   \otimes \left| \psi  \right\rangle _{A_N N}^ -  } \right\}\left| \chi  \right\rangle _{(N + 1)(N + 2)......(2N - 1)2N}^{(4^N)}  \\ 
 \end{array} \right]
\end{eqnarray}
\end{widetext}
where $\left| \chi  \right\rangle _{(N + 1)(N + 2)......(2N - 1)2N}^{(1)}$ $-$ $\left| \chi  \right\rangle _{(N + 1)(N + 2)......(2N - 1)2N}^{(4^N)}$ are the states of the receiver's qubits depending on the measurement outcomes of all the senders and contains all the information about the senders' individual qubits. 
 If a six-qubit entangled resource is used then the $\left|\chi\right\rangle_{456}^{(1)}$ state corresponding to the measurement outcomes $\left|\phi\right\rangle_{A_{1}1}^{+}$, $\left|\phi\right\rangle_{A_{2}2}^{+}$, and $\left|\phi\right\rangle_{A_{3}3}^{+}$ can be written as
\begin{widetext}
\begin{eqnarray}
\left| \chi  \right\rangle _{456}^{(1)}  &=& \left[a_{A_1} a_{A_2 } a_{A_3 } \left| {000} \right\rangle _{456}  + a_{A_1 } a_{A_2 } b_{A_3 } \left| {001} \right\rangle _{456} + a_{A_1 } b_{A_2 } a_{A_3 } \left| {011} \right\rangle _{456}+ a_{A_1 } b_{A_2 } b_{A_3 } \left| {010} \right\rangle _{456}\right. \nonumber \\  &+& \left. b_{A_1 } a_{A_2 } a_{A_3 } \left| {101} \right\rangle _{456}  + b_{A_1 } a_{A_2 } b_{A_3 } \left| {100} \right\rangle _{456}  + b_{A_1 } b_{A_2 } a_{A_3 } \left| {110} \right\rangle _{456}  + b_{A_1 } b_{A_2 } b_{A_3 } \left| {111} \right\rangle _{456} \right]
\end{eqnarray}
\end{widetext}
As discussed above, the receiver would not be able to separate the exact votes sent by the senders as the joint state of receiver's qubit cannot be written as the direct product state of the individual qubits. However, if we apply two CNOT operations with qubits 4 and 5 as controls and qubit 6 as target then the joint state of receiver's qubits can be re-expressed as 

\begin{widetext}
\begin{eqnarray}
\left| \chi  \right\rangle _{456}^{(1)}  &=& \left[a_{A_1} a_{A_2 } a_{A_3 } \left| {000} \right\rangle _{456}  + a_{A_1 } a_{A_2 } b_{A_3 } \left| {001} \right\rangle _{456} + a_{A_1 } b_{A_2 } a_{A_3 } \left| {010} \right\rangle _{456}+ a_{A_1 } b_{A_2 } b_{A_3 } \left| {011} \right\rangle _{456}\right. \nonumber \\  &+& \left. b_{A_1 } a_{A_2 } a_{A_3 } \left| {100} \right\rangle _{456}  + b_{A_1 } a_{A_2 } b_{A_3 } \left| {101} \right\rangle _{456}  + b_{A_1 } b_{A_2 } a_{A_3 } \left| {110} \right\rangle _{456}  + b_{A_1 } b_{A_2 } b_{A_3 } \left| {111} \right\rangle _{456} \right] \nonumber \\
&=& \left[a_{A_{1}} \left| 0 \right\rangle_{4}   + b_{A_{1}} \left| 1_{4}  \right\rangle\right] \otimes \left[a_{A_{2}} \left| 0 \right\rangle_{5}   + b_{A_{2}} \left| 1_{5}  \right\rangle\right] \otimes \left[a_{A_{3}} \left| 0 \right\rangle_{6}   + b_{A_{3}} \left| 1_{6}  \right\rangle\right],
\end{eqnarray}
\end{widetext}
respectively. Thus, the receiver can successfully separate the exact qubits teleported by the senders. However, for all the other measurement outcomes of the senders, the receiver would need to perform single qubit unitary operations in addition to the CNOT transformations. For example, if a six-qubit quantum channel is used then for the measurement outcomes $\left|\phi\right\rangle_{A_{1}1}^{+}$, $\left|\phi\right\rangle_{A_{2}2}^{+}$, and $\left|\psi\right\rangle_{A_{3}3}^{+}$ of the senders, the receiver needs to apply a $\sigma_{x}^{6}$ operation on the qubit 6. The protocol is successful for all the measurement outcomes of the senders. The required single-qubit unitary transformations for the use of six-qubit entangled channels is summarized in Table 2. For this protocol to be successful the receiver must
know the distribution of entangled qubits so that he/she can apply the correct unitary transformations to recover
the qubits exactly. However, this does not allow the receiver to gain any information about the individual qubits of the senders.

\newpage

\begin{table*}
\caption{\label{tab:table3} 3-qubit teleportation: unitary transformations required for the receiver to recover the 3-qubit teleported states depending on the measurement outcomes of the three senders.}
\begin{ruledtabular}
\begin{tabular}{ccccc}
First  & Second   & Third   & Final state with the receiver & Unitary transformations  \\ sender's & sender's & sender's & after the CNOT operations & required to obtain the \\ measurement & measurement & measurement & & correct votes corresponding  \\ outcomes & outcomes & outcomes & & to the measurement outcomes \\ & & & & $+++$, $++-$, $+-+$, $+--$ \\ & & & & $-++$, $-+-$, $--+$, $---$ \\ \hline
$\left|\phi\right\rangle_{A_{1}1}^{\pm}$&$\left|\phi\right\rangle_{A_{2}2}^{\pm}$ & $\left|\phi\right\rangle_{A_{3}3}^{\pm}$ & $\left[a_{A_{1}1}\left|0\right\rangle \pm b_{A_{1}1}\left|1\right\rangle\right]_{4}\otimes\left[a_{A_{2}2}\left|0\right\rangle \pm b_{A_{2}2}\left|1\right\rangle\right]_{5}$ & $I$, $\sigma_{z}^{6}$, $\sigma_{z}^{5}$,  $\sigma_{z}^{5}\sigma_{z}^{6}$,  \\ & & & $\otimes \left[a_{A_{3}3}\left|0\right\rangle \pm b_{A_{3}3}\left|1\right\rangle\right]_{6}$ & $\sigma_{z}^{4}$, $\sigma_{z}^{4}\sigma_{z}^{6}$, $\sigma_{z}^{4}\sigma_{z}^{5}$, $\sigma_{z}^{4}\sigma_{z}^{5}\sigma_{z}^{6}$ \\ & & & & \\

$\left|\phi\right\rangle_{A_{1}1}^{\pm}$&$\left|\phi\right\rangle_{A_{2}2}^{\pm}$ & $\left|\psi\right\rangle_{A_{3}3}^{\pm}$ & $\left[a_{A_{1}1}\left|0\right\rangle \pm b_{A_{1}1}\left|1\right\rangle\right]_{4}\otimes\left[a_{A_{2}2}\left|0\right\rangle \pm b_{A_{2}2}\left|1\right\rangle\right]_{5}$ & $\sigma_{x}^{6}$, $\sigma_{y}^{6}$, $\sigma_{z}^{5}\sigma_{x}^{6}$,  $\sigma_{z}^{5}\sigma_{x}^{6}$  \\ & & & $\otimes \left[a_{A_{3}3}\left|1\right\rangle \pm b_{A_{3}3}\left|0\right\rangle\right]_{6}$ & $\sigma_{z}^{4}\sigma_{x}^{6}$, $\sigma_{z}^{4}\sigma_{y}^{6}$, $\sigma_{z}^{4}\sigma_{z}^{5}\sigma_{x}^{6}$,  $\sigma_{z}^{4}\sigma_{z}^{5}\sigma_{x}^{6}$ \\ & & & & \\

$\left|\phi\right\rangle_{A_{1}1}^{\pm}$&$\left|\psi\right\rangle_{A_{2}2}^{\pm}$ & $\left|\phi\right\rangle_{A_{3}3}^{\pm}$ & $\left[a_{A_{1}1}\left|0\right\rangle \pm b_{A_{1}1}\left|1\right\rangle\right]_{4}\otimes\left[a_{A_{2}2}\left|1\right\rangle \pm b_{A_{2}2}\left|0\right\rangle\right]_{5}$ & $\sigma_{x}^{5}$, $\sigma_{x}^{5}\sigma_{z}^{6}$, $\sigma_{y}^{5}$, $\sigma_{y}^{5}\sigma_{z}^{6}$  \\ & & & $\otimes \left[a_{A_{3}3}\left|0\right\rangle \pm b_{A_{3}3}\left|1\right\rangle\right]_{6}$ & $\sigma_{z}^{4}\sigma_{x}^{5}$, $\sigma_{z}^{4}\sigma_{x}^{5}\sigma_{z}^{6}$, $\sigma_{z}^{4}\sigma_{y}^{5}$, $\sigma_{z}^{4}\sigma_{y}^{5}\sigma_{z}^{6}$ \\ & & & & \\
 
$\left|\phi\right\rangle_{A_{1}1}^{\pm}$&$\left|\psi\right\rangle_{A_{2}2}^{\pm}$ & $\left|\psi\right\rangle_{A_{3}3}^{\pm}$ & $\left[a_{A_{1}1}\left|0\right\rangle \pm b_{A_{1}1}\left|1\right\rangle\right]_{4}\otimes\left[a_{A_{2}2}\left|1\right\rangle \pm b_{A_{2}2}\left|0\right\rangle\right]_{5}$ & $\sigma_{x}^{5}\sigma_{x}^{6}$, $\sigma_{x}^{5}\sigma_{y}^{6}$, $\sigma_{y}^{5}\sigma_{x}^{6}$, $\sigma_{y}^{5}\sigma_{y}^{6}$  \\ & & & $\otimes \left[a_{A_{3}3}\left|1\right\rangle \pm b_{A_{3}3}\left|0\right\rangle\right]_{6}$ & $\sigma_{z}^{4}\sigma_{x}^{5}\sigma_{x}^{6}$, $\sigma_{z}^{4}\sigma_{x}^{5}\sigma_{y}^{6}$, $\sigma_{z}^{4}\sigma_{y}^{5}\sigma_{x}^{6}$, $\sigma_{z}^{4}\sigma_{y}^{5}\sigma_{y}^{6}$ \\ & & & & \\

$\left|\psi\right\rangle_{A_{1}1}^{\pm}$&$\left|\phi\right\rangle_{A_{2}2}^{\pm}$ & $\left|\phi\right\rangle_{m_{3}3}^{\pm}$ & $\left[a_{A_{1}1}\left|1\right\rangle \pm b_{A_{1}1}\left|0\right\rangle\right]_{4}\otimes\left[a_{A_{2}2}\left|0\right\rangle \pm b_{A_{2}2}\left|1\right\rangle\right]_{5}$ & $\sigma_{x}^{4}$, $\sigma_{x}^{4}\sigma_{z}^{6}$, $\sigma_{x}^{4}\sigma_{z}^{5}$,  $\sigma_{x}^{4}\sigma_{z}^{5}\sigma_{z}^{6}$  \\ & & & $\otimes \left[a_{A_{3}3}\left|0\right\rangle \pm b_{A_{3}3}\left|1\right\rangle\right]_{6}$ & $\sigma_{y}^{4}$, $\sigma_{y}^{4}\sigma_{z}^{6}$, $\sigma_{y}^{4}\sigma_{z}^{5}$,  $\sigma_{y}^{4}\sigma_{z}^{5}\sigma_{z}^{6}$ \\ & & & & \\
 
$\left|\psi\right\rangle_{A_{1}1}^{\pm}$&$\left|\phi\right\rangle_{A_{2}2}^{\pm}$ & $\left|\psi\right\rangle_{A_{3}3}^{\pm}$ & $\left[a_{A_{1}1}\left|1\right\rangle \pm b_{A_{1}1}\left|0\right\rangle\right]_{4}\otimes\left[a_{A_{2}2}\left|0\right\rangle \pm b_{A_{2}2}\left|1\right\rangle\right]_{5}$ & $\sigma_{x}^{4}\sigma_{x}^{6}$, $\sigma_{x}^{4}\sigma_{y}^{6}$, $\sigma_{x}^{4}\sigma_{z}^{5}\sigma_{x}^{6}$,  $\sigma_{x}^{4}\sigma_{z}^{5}\sigma_{y}^{6}$  \\ & & & $\otimes \left[a_{A_{3}3}\left|1\right\rangle \pm b_{A_{3}3}\left|0\right\rangle\right]_{6}$ & $\sigma_{y}^{4}\sigma_{x}^{6}$, $\sigma_{y}^{4}\sigma_{y}^{6}$, $\sigma_{y}^{4}\sigma_{z}^{5}\sigma_{x}^{6}$,  $\sigma_{y}^{4}\sigma_{z}^{5}\sigma_{y}^{6}$ \\ & & & & \\

$\left|\psi\right\rangle_{A_{1}1}^{\pm}$&$\left|\psi\right\rangle_{A_{2}2}^{\pm}$ & $\left|\phi\right\rangle_{A_{3}3}^{\pm}$ & $\left[a_{A_{1}1}\left|1\right\rangle \pm b_{A_{1}1}\left|0\right\rangle\right]_{4}\otimes\left[a_{A_{2}2}\left|1\right\rangle \pm b_{A_{2}2}\left|0\right\rangle\right]_{5}$ & $\sigma_{x}^{4}\sigma_{x}^{5}$, $\sigma_{x}^{4}\sigma_{y}^{5}$, $\sigma_{x}^{4}\sigma_{x}^{5}\sigma_{z}^{6}$,  $\sigma_{x}^{4}\sigma_{y}^{5}\sigma_{z}^{6}$   \\ & & & $\otimes \left[a_{A_{3}3}\left|0\right\rangle \pm b_{A_{3}3}\left|1\right\rangle\right]_{6}$ & $\sigma_{y}^{4}\sigma_{x}^{5}$, $\sigma_{y}^{4}\sigma_{y}^{5}$, $\sigma_{y}^{4}\sigma_{x}^{5}\sigma_{z}^{6}$,  $\sigma_{y}^{4}\sigma_{y}^{5}\sigma_{z}^{6}$ \\ & & & & \\

$\left|\psi\right\rangle_{A_{1}1}^{\pm}$&$\left|\psi\right\rangle_{A_{2}2}^{\pm}$ & $\left|\psi\right\rangle_{A_{3}3}^{\pm}$ & $\left[a_{A_{1}1}\left|1\right\rangle \pm b_{A_{1}1}\left|0\right\rangle\right]_{4}\otimes\left[a_{A_{2}2}\left|1\right\rangle \pm b_{A_{2}2}\left|0\right\rangle\right]_{5}$ & $\sigma_{x}^{4}\sigma_{x}^{5}\sigma_{x}^{6}$, $\sigma_{x}^{4}\sigma_{x}^{5}\sigma_{y}^{6}$, $\sigma_{x}^{4}\sigma_{y}^{5}\sigma_{x}^{6}$,  $\sigma_{x}^{4}\sigma_{y}^{5}\sigma_{y}^{6}$ \\ & & & $\otimes \left[a_{A_{3}3}\left|1\right\rangle \pm b_{A_{3}3}\left|0\right\rangle\right]_{6}$ & $\sigma_{y}^{4}\sigma_{x}^{5}\sigma_{x}^{6}$, $\sigma_{y}^{4}\sigma_{x}^{5}\sigma_{y}^{6}$, $\sigma_{y}^{4}\sigma_{y}^{5}\sigma_{x}^{6}$,  $\sigma_{y}^{4}\sigma_{y}^{5}\sigma_{y}^{6}$ \\ 
\end{tabular}
\end{ruledtabular}
\end{table*}

\end{document}